\documentstyle[twoside,fleqn,epsf,espcrc2]{article}
\begin{document}
\let\addressmark\relax

\renewcommand{\thefootnote}{\fnsymbol{footnote}}

\title{\vglue-.7in
\font\fortssbx=cmssbx10 scaled \magstep1
\hbox to \hsize{%
\hbox{\fortssbx University of Wisconsin - Madison}
\hfill\vtop{\normalsize\hbox{\bf MADPH-98-1078}
                \hbox{September 1998}
                \hbox{\hfil}}}
The AMANDA Neutrino Telescope\footnotemark}

\author{Francis Halzen, for the AMANDA Collaboration$^\dagger$\address{Physics Department, University of Wisconsin, Madison, WI 53706, USA}}

\begin{abstract}
With an effective telescope area of order $10^4$~m$^2$ for TeV neutrinos, a threshold near $\sim$50~GeV and a pointing accuracy of 2.5~degrees per muon track, the AMANDA detector represents the first of a new generation of high energy neutrino telescopes, reaching a scale envisaged over 25 years ago. We describe early results on the calibration of natural deep ice as a particle detector as well as on AMANDA's performance as a neutrino telescope.\end{abstract}

\maketitle
\thispagestyle{empty}

\footnotetext{Talk presented at the {\it 18th International Conference on Neutrino Physics and Astrophysics (Neutrino\,98)}, Takayama, Japan, June 1998.}

\renewcommand{\topfraction}{1.0}    
\renewcommand{\bottomfraction}{1.0}
\renewcommand{\textfraction}{0.0}

\section{\uppercase{Introduction and Summary}}

The Antarctic Muon and Neutrino Detector Array AMANDA is a  multi-purpose instrument; its science missions cover particle physics, astronomy and astrophysics, cosmology and
cosmic ray physics\cite{pr}.  Its deployment creates new opportunities for glaciology\cite{science}. The first-generation detector is designed to reach a relatively large telescope area and detection volume for a neutrino threshold not higher than 100\,GeV. This relatively low threshold permits calibration of the novel instrument on the known flux of atmospheric neutrinos.  Its architecture has been optimized for reconstructing the Cherenkov light front radiated by up-going, neutrino-induced muons which must be identified in a background of down-going, cosmic ray muons which are more than $10^5$ times more frequent for a depth of 1--2\,kilometer.

The status of the AMANDA project can be summarized as follows:

\begin{itemize}

\item
Construction of the first generation\break AMANDA detector\cite{barwick} was completed in the austral summer 96--97. It consists of 300 optical modules deployed at a depth of 1500--2000~m; see Fig.\,1. An optical module (OM) consists of an 8\,inch photomultiplier tube and nothing else. OM's have only failed during deployment, at a rate of less than 3 percent.

\item
Data taken with 80 OM's, deployed one year earlier in order to verify the optical properties
of the deep ice, have been analysed. We will present the results here. This partially deployed detector will be referred to as AMANDA-B4. Reconstructed up-going muons are found at a rate consistent with the expected flux of atmospheric neutrinos. The exercise shows that calibration of the full detector on atmospheric neutrinos of approximately 100\,GeV energy and above, is possible as we will show further~on.  

\item
First calibration of the full detector is now completed and analysis of the first year of data is in progress. Preliminary results based on the analysis of 1 month of data confirm the performance of the detector derived from the analysis of AMANDA-B4 data. Events reconstructed as going  upwards, like the one shown in Fig.\,2, are found, as expected.   

\end{itemize}

\begin{figure*}
\epsfxsize=12cm\epsffile{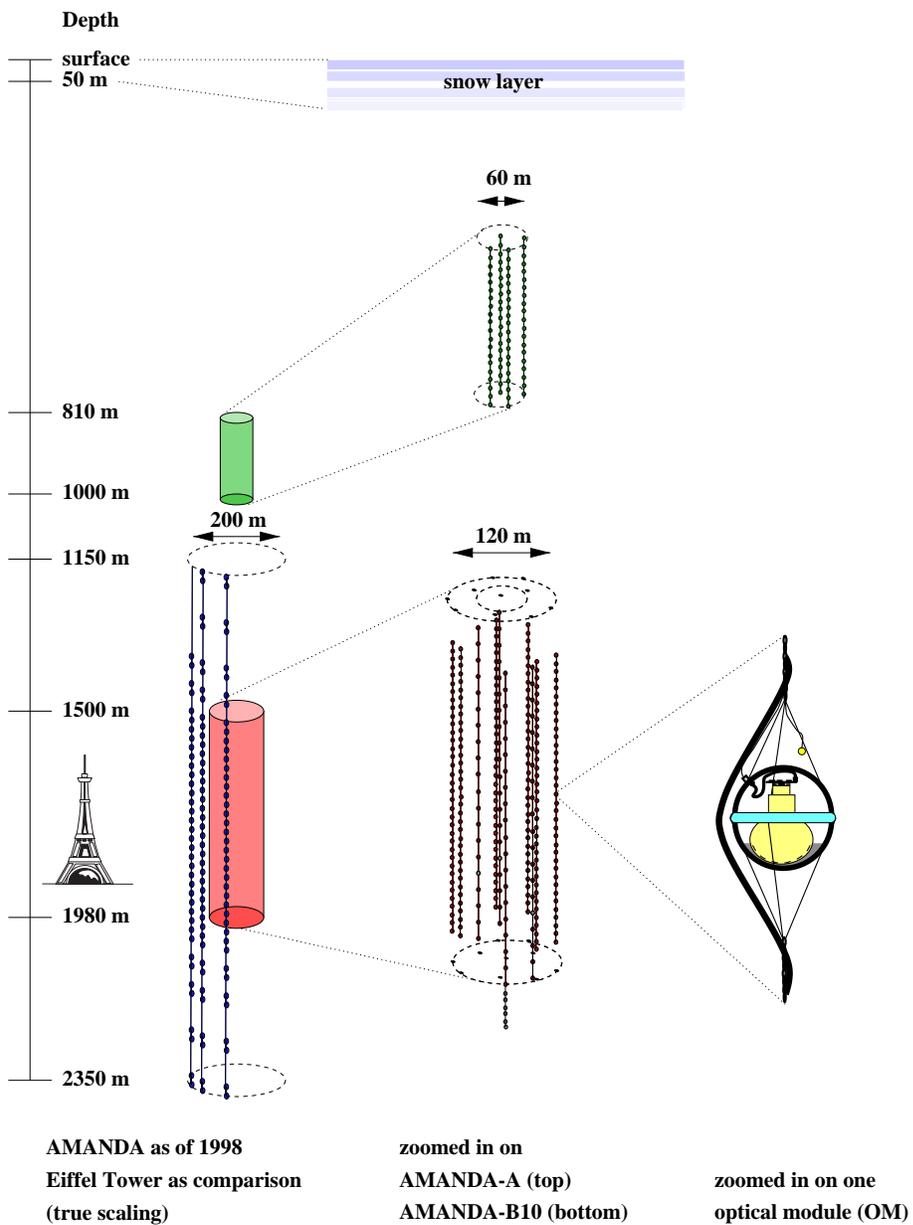}

\caption{The Antarctic Muon And Neutrino Detector Array (AMANDA).}
\end{figure*}

\begin{figure*}
\epsfxsize=12cm\epsffile{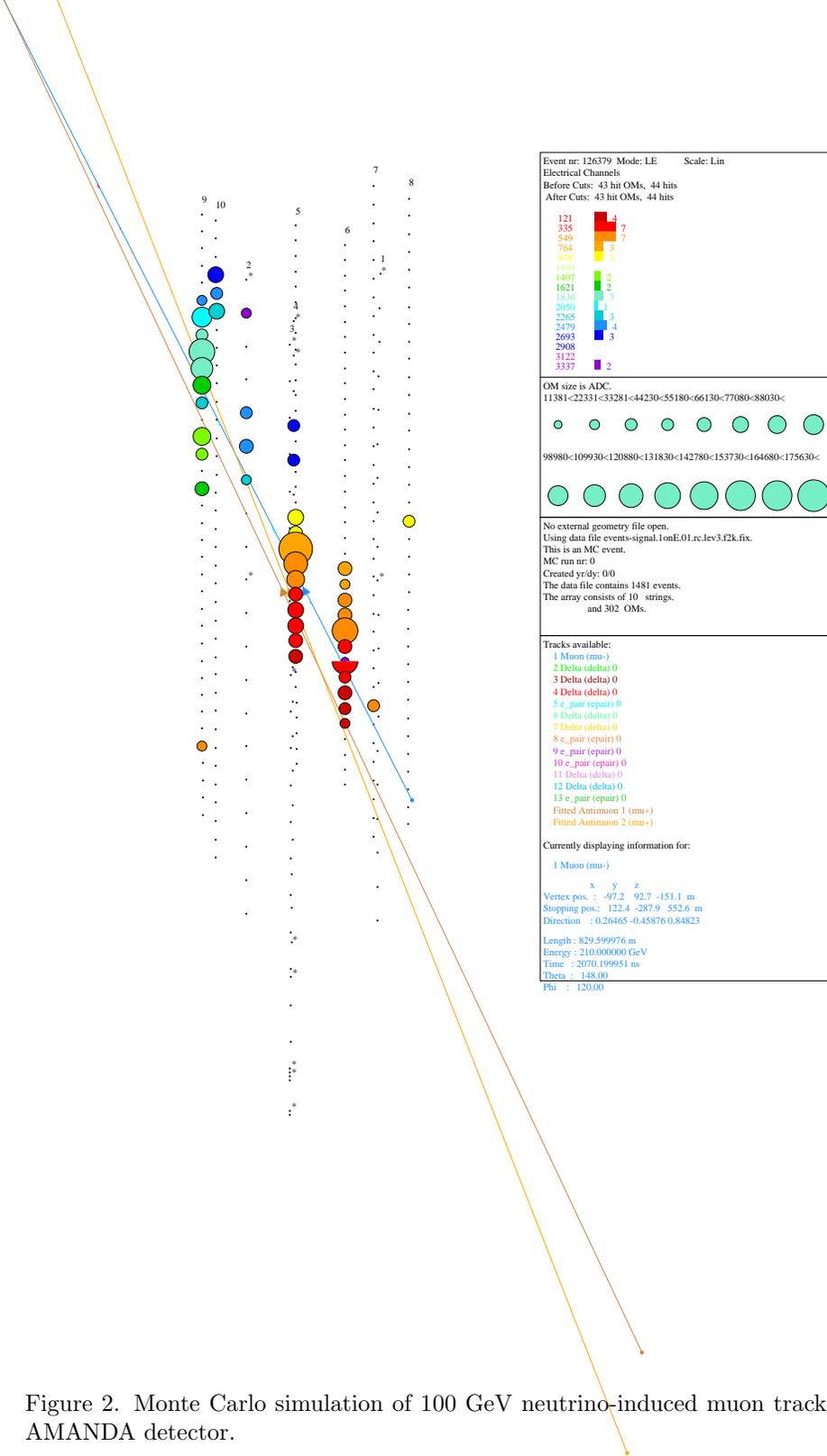}

\caption{Monte Carlo simulation of 100 GeV neutrino-induced muon track recorded in the completed AMANDA detector.}
\end{figure*}

\begin{figure}[t]
\centering\leavevmode
\epsfxsize=7.5cm\epsffile{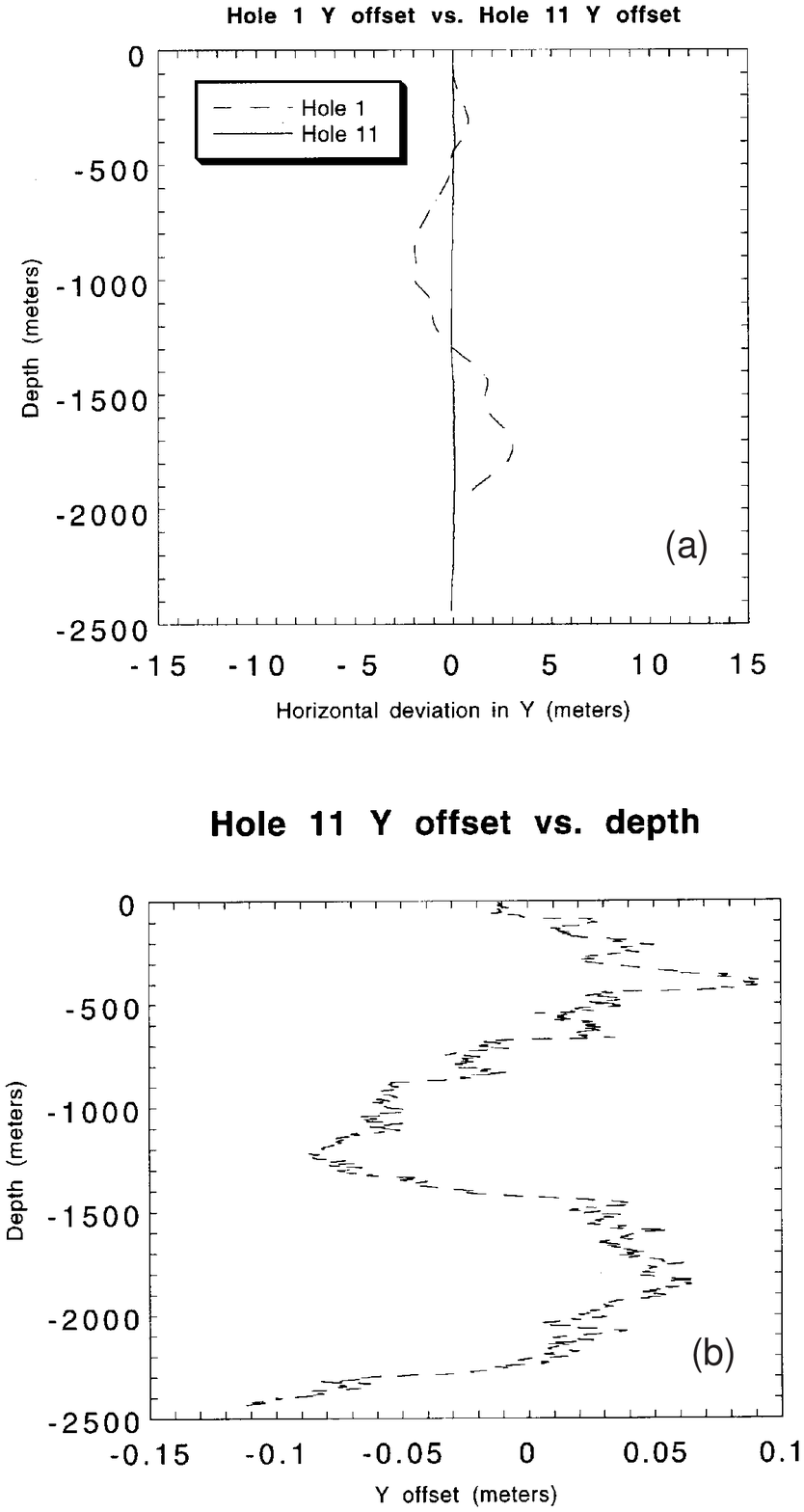}

\caption{a) Progress in drilling: telemetry data from the drill compare the deviation from vertical. b) Excursions transverse to the vertical direction are smaller than 1\,m over 2.4\,km.}
\end{figure}

As part of a research and development effort preparatory to developing a kilometer-scale neutrino detector, we have deployed 3 strings, instrumented with 42 OMs between 1.3 and 2.4 kilometers; see Fig.\,1. The strings deviated from vertical by less than 1\,m over 2.4\,km; see Fig.\,3. They also form part of  an intermediate detector, AMANDA\,II,  which will extend the present telescope by approximately an order of magnitude in affective area for TeV energies. It will be completed in 99-00 with the addition of eight more strings. The analogue signals made by photoelectron pulses in the new OMs are transferred to the surface over both twisted pair and fiber optic cables. The relative sharpness of the pulses at the surface is compared in Fig.\,4. Also, bright light sources surrounding a pair of TV cameras were lowered into the last hole. The resulting images visually confirm the exceptional clarity of the ice inferred from previous indirect measurements.  

After a brief review of our results on the optics of the ice, we will discuss muon track reconstruction and the status of the calibration of the detector on the flux of atmospheric neutrinos. We will conclude with a brief description of the data analysis of the first year of a data taken with the completed detector.

\begin{figure}[t]
\centering\leavevmode
\epsfxsize=7.5cm\epsffile{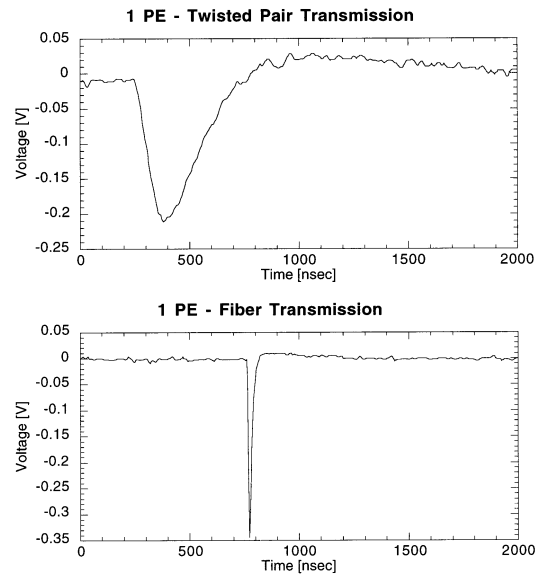}

\caption{Time profile of {\it the same} single photoelectron pulse after transmission over a twisted pair (top) and fiber optic cable (bottom). The signal is from an optical module deployed at a depth of 1.7\,km
for strings 1 and 13.}
\end{figure}

\section{\uppercase{Optics of Deep Ice}}

\looseness=-1
As anticipated from transparency measurements performed with shallow strings above 1\,km depth\cite{science} (see Fig.\,1), ice is bubble-free at 1400--1500~meters and below. The performance of the AMANDA detector is encapsulated in the event shown in Fig.\,5. Coincident events between AMANDA-B4 and the four shallow strings have been triggered at a rate of 0.1\,Hz. Every 10 seconds a cosmic ray muon is tracked over 1.2 kilometers. The contrast in detector response between the strings near 1 and 2\,km depths is striking: while the Cherenkov photons diffuse on remnant bubbles in the shallow ice, a straight track with velocity $c$ is registered in the deeper ice. The optical quality of the deep ice can be assessed by viewing the OM signals from a single muon triggering 2 strings separated by 79.5\,m; see Fig.\,5b. The separation of the photons along the Cherenkov cone is well over 100\,m, yet, despite some evidence of scattering, the speed-of-light propagation of the track can be readily identified.

\renewcommand{\thefigure}{\arabic{figure}a}
\begin{figure}[t]
\centering\leavevmode
\epsfxsize=7.5cm\epsffile{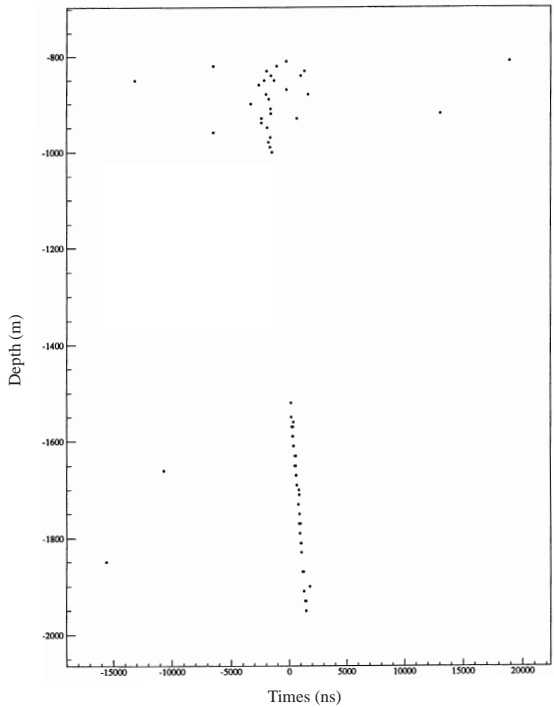}

\caption{Cosmic ray muon track triggered by both shallow and deep AMANDA
OM's. Trigger times of the optical modules are shown as a function of depth.
The diagram shows the diffusion of the track by bubbles above 1~km depth.
Early and late hits, not associated with the track, are photomultiplier noise.}
\end{figure}

\addtocounter{figure}{-1}\renewcommand{\thefigure}{\arabic{figure}b}
\begin{figure}[t]
\centering\leavevmode
\epsfxsize=7.5cm\epsffile{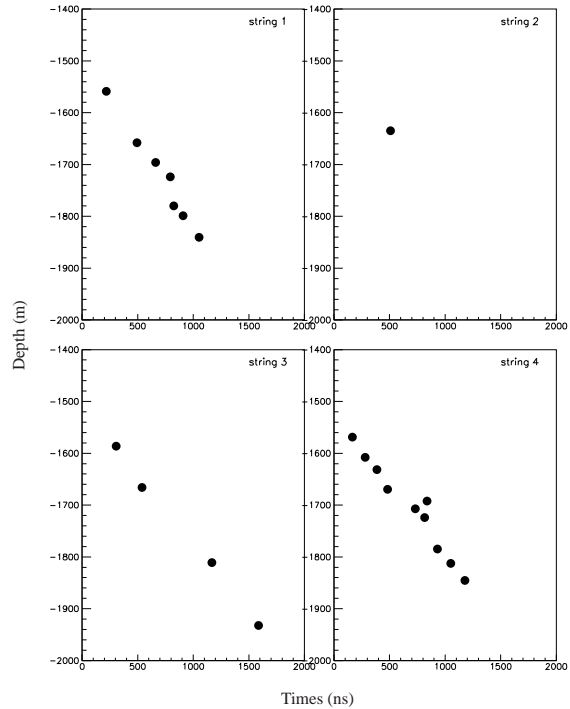}

\caption{Cosmic ray muon track triggered by both shallow and deep AMANDA
OM's. Trigger times are shown separately for each string in the deep detector.
In this event the muon mostly triggers OM's on strings 1 and 2, which are
separated by 77.5~m. }
\end{figure}
\renewcommand{\thefigure}{\arabic{figure}}

The optical properties of the ice are quantified by studying the propagation in the ice of pulses of laser light of nanosecond duration. The arrival times of the photons after 20~m and 40~m are shown in Fig.\,6 for the shallow and deep ice\cite{serap}. The distributions have been normalized to equal areas; in reality, the probability that a photon travels 70~m in the deep ice is ${\sim}10^7$ times larger. These critical results have been verified by the deployment of nitrogen lasers, pulsed LED's and DC lamps in the deep ice; see Table~1. We have established that ice is an adequate medium to do neutrino astronomy. A comparison of the optical properties of ice, lake and ocean detectors is summarized in Table~2.

\begin{figure*}[t]
\centering\leavevmode
\epsfxsize=15cm\epsffile{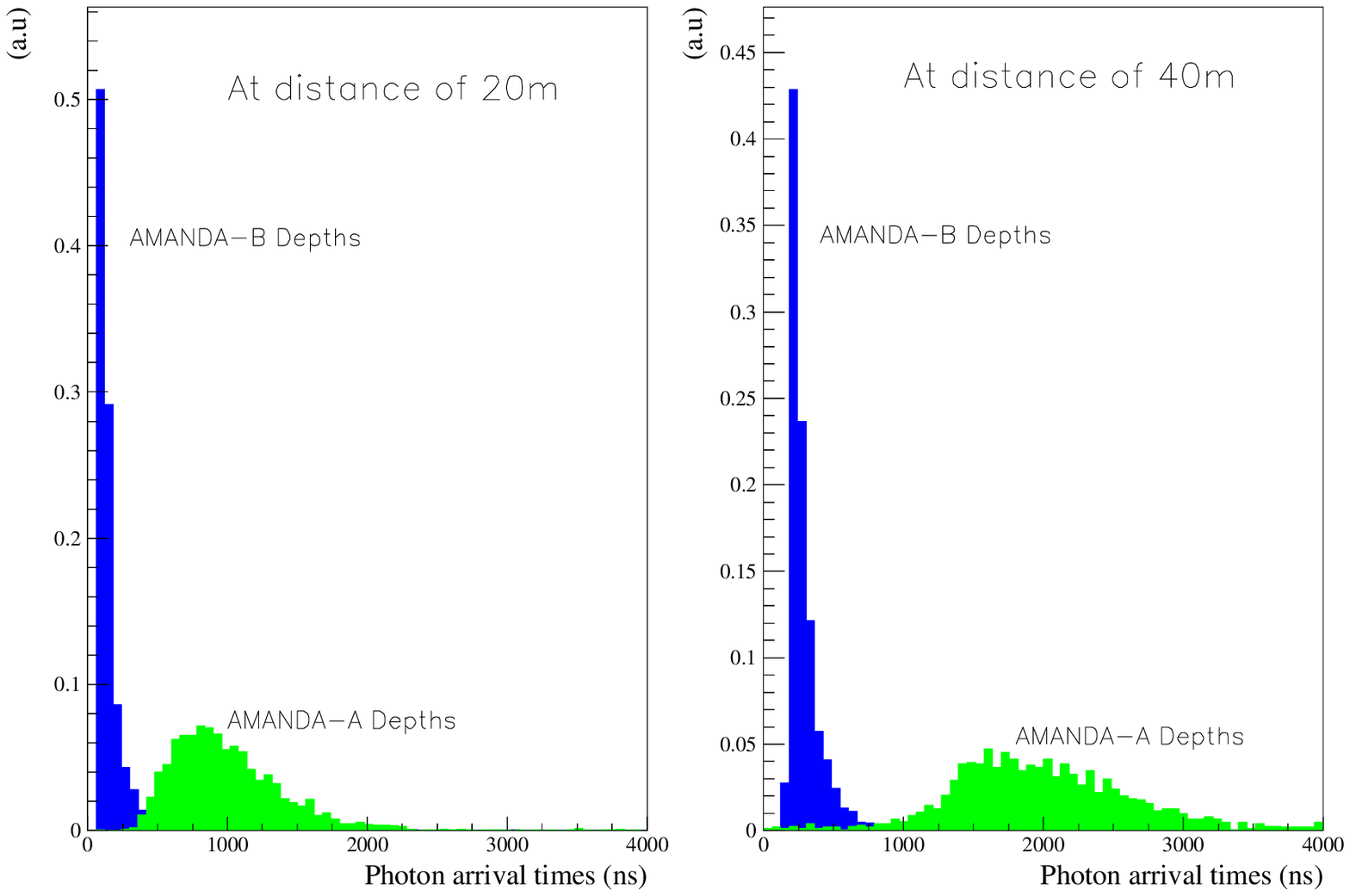}

\caption{Propagation of 510~nm photons indicate bubble-free ice below 1500~m,
in contrast to ice with some remnant bubbles above 1.4~km.}
\end{figure*}


Inter-string laser shots are also used to determine the geometry of the detector. In conjunction with telemetry from the drill, the OMs have been positioned with an absolute precision of better than 1~meter. Mapping the detector has been by far the most challenging aspect of the calibration of this novel instrument. A precise knowledge of the location of the optical sensors is crucial for track reconstruction. Therefore, two completely independent methods were developed for the final determination of the geometry. One method makes use of drill data. A variety of sensors are installed in the drill to determine its speed and direction during drilling. Every second a data string is transmitted to the control system  
and recorded. The analysis of this data provides the first information about the string position. The depth of the string is independently determined with pressure sensors. The final positioning of the strings is done with a laser calibration system. Laser pulses (532~nm) are transmitted with optical fibers to every optical module on strings 1--4, and to every second module on strings 5--10. After the timing calibration is completed, the laser calibration provides time of flight measurements to determine the distances between strings and a check on possible vertical offsets. More than a hundred laser runs provide a large data base, both to determine the geometry and to verify the timing calibration. Figure~7 shows laser data from string 8 recorded on string 7, with the results from a global fit to data from all 10 strings plotted as a solid line. The vertical offset between the strings from pressure sensor data was found to be 0.9~m and the distance between them has been determined to 29.9~m. The errors given in the figure are the statistical errors from the global fit. The position error of the optical sensors is less than 1~m, thus matching the time resolution of the sensors.

\begin{table}[t]
\def\arraystretch{1.5}\tabcolsep=0em
\caption{Complementary tools used in the determination of the optical properties of in-situ South Pole ice.}
\smallskip
\centering\leavevmode
\begin{tabular}{@{$\bullet$\ }l}
\hline\noalign{\vskip.3ex}
\parbox[t]{7cm}{surface YAG laser (410--600 nm) connected by fiber optic to $\sim300$ diffuser balls}\\
5 N$_2$ lasers (337 nm) between 1300--2300 m\\
pulsed LEDs (390, 450 nm)\\
DC lamps\\
DC beacons\\
multiple radio antennas (150--300 m)\\
2 TV cameras to 2400 m\\[.3ex]
\hline
\end{tabular}
\end{table}

\begin{table}[t]
\def\arraystretch{1.5}\tabcolsep=.2em
\caption{Optical properties of South Pole ice at 1750\,m, Lake Baikal water at 1\,km, and the range of results from measurements in ocean water below 4\,km.}
\smallskip
\centering\leavevmode
\begin{tabular}{lccc}
\hline
& (1700 m)&&\\[-1.5ex]
 & AMANDA& BAIKAL& OCEAN\\
\hline
attenuation& $\sim 30$ m& $\sim 20$ m& 50--55 m\\
absorption& $105\pm10$ m& 20 m& ---\\
(refers to& 335--400 nm& 470 nm$^*$& 470 nm$^*$\\[-2ex]
peak value)&&&\\
scattering& $24\pm 2$ m& 150--300 m& ---\\[-2ex]
length&&&\\
\hline
\multicolumn{4}{l}{$^*$smaller for bluer wavelengths}
\end{tabular}
\end{table}

\begin{figure}[t]
\epsfxsize=7cm\epsffile{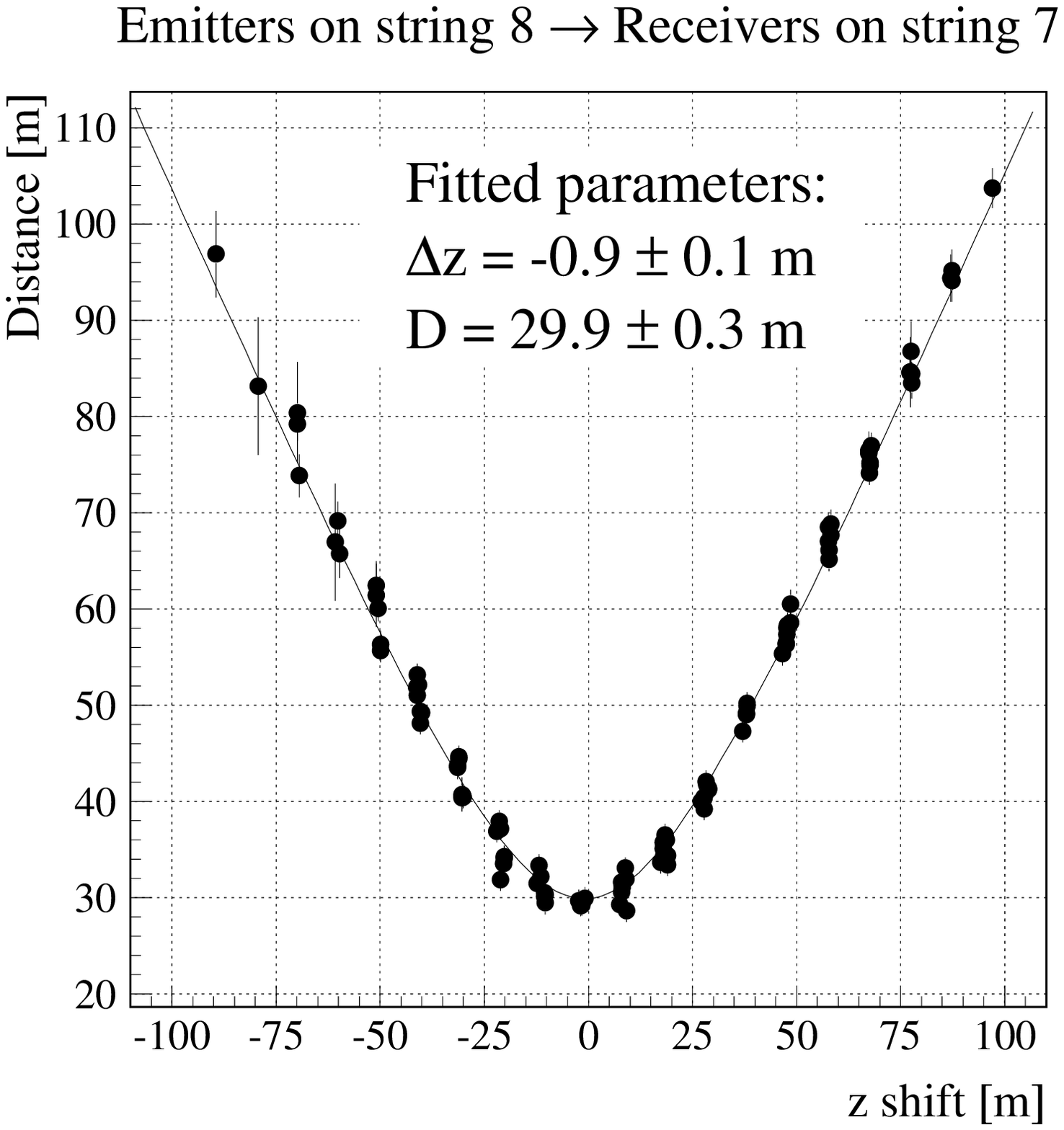}

\caption{Shots of laserlight determine the distance and the vertical offset between strings.}
\end{figure}

\section{\uppercase{Reconstruction of Muon\hfil\break Tracks}}

The AMANDA detector was antecedently proposed on the premise that inferior properties of ice as a particle detector with respect to water could be compensated by additional optical modules. The technique was supposed to be a factor $5 {\sim} 10$ more cost-effective and, therefore, competitive. The design was based on then current information\cite{dublin} that the absorption length at 370~nm, the wavelength where photomultipliers are maximally efficient, had been measured to be 8~m. The strategy would have been to use a large number of closely spaced OM's to overcome the short absorption length. Muon tracks triggering 6 or more OM's were reconstructed with degree accuracy. Taking data with a simple majority trigger of 6 OM's or more, at 100\,Hz yielded an average effective area of $10^4$~m$^2$, somewhat smaller for atmospheric neutrinos and significantly larger for the high energy signals.

The reality is that the absorption length is 100\,m or more, depending on depth\cite{science}. With such a large absorption length, scattering becomes a critical issue. The scattering length is 25--30~m (preliminary; this number represents an average value which may include the combined effects of deep ice and the refrozen ice disturbed by the hot water drilling). Because of the large absorption length, OM spacings are now similar, actually larger, than those of proposed water detectors. A typical event triggers 20 OM's, not~6. Of these more than 5 photons are, on average, ``not scattered\rlap". They are referred to as direct photons, i.e.\ photons which arrive within time residuals of $[-15; 25]$\,ns relative to the calculated time it takes for unscattered Cherenkov photons to reach the OM from the reconstructed muon track. The choice of residual reflects the present resolution of our time measurements and allows for delays of slightly scattered photons. In the end, reconstruction is therefore as before, although additional information can be extracted from scattered photons by minimizing a likelihood function which matches their observed and expected delays\cite{christopher}.

The method is illustrated with AMANDA-B4 data in Fig.\,8, where the measured arrival directions of background cosmic ray muon tracks, reconstructed with 5 
or more unscattered photons, are confronted with their known angular distribution. There is an additional cut in Fig.\,8 which requires that the track, reconstructed from timing information, actually traces the spatial positions of the OM's in the trigger. The power of this cut, especially for events recorded with only 4 strings, is very revealing. In a kilometer-scale detector, geometrical track reconstruction using only the positions of triggered OM's is sufficient to achieve degree accuracy in zenith angle. We conclude from Fig.\,8 that the agreement between data and Monte Carlo simulation is adequate. Less than one in $10^5$ tracks is misreconstructed as originating below the detector\cite{serap}.

\begin{figure*}
\centering\leavevmode
\epsfxsize=14cm\epsffile{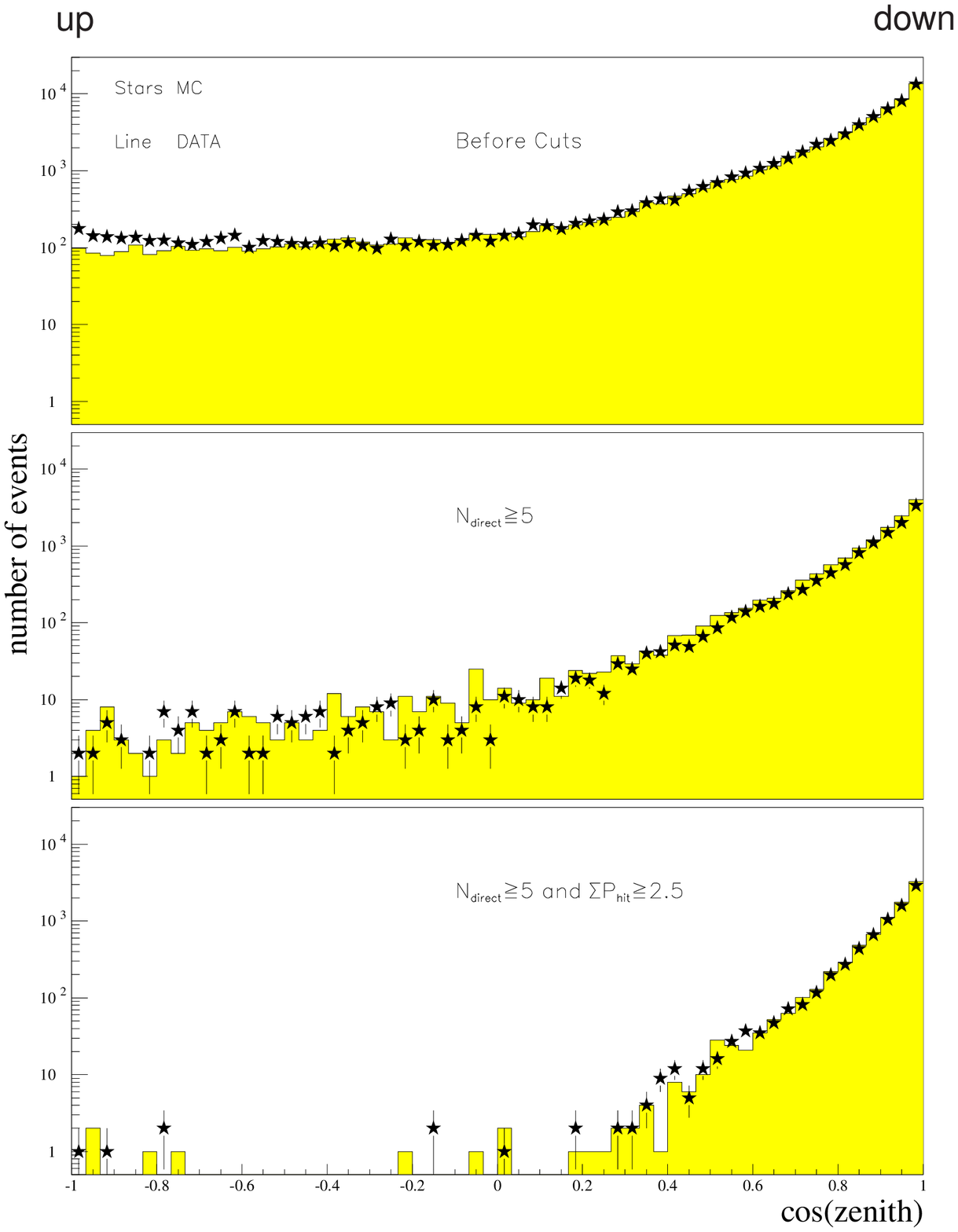}

\caption{Reconstructed zenith angle distribution of muons: data and Monte Carlo. The relative normalization has not been adjusted at any level. The plot demonstrates a rejection of cosmic ray muons at a level of 10$^{-5}$ with only 80 OMs.}
\end{figure*}

Visual inspection reveals that the misreconstructed tracks are mostly showers, radiated by muons or initiated by electron neutrinos, which are misreconstructed as up-going tracks of muon neutrino origin. They can be readily identified on the basis of the characteristic nearly isotropic distribution of the OM amplitudes, and by the fact that the direct hits occur over a short distance near the origin of the shower, rather than spread over a longer muon~track.

We have verified the angular resolution of AMANDA-B4 by reconstructing muon tracks registered in coincidence with a surface air shower array SPASE\cite{miller}. Figure\,9 demonstrates that the zenith angle distribution of the coincident SPASE-AMANDA cosmic ray beam reconstructed by the surface array is quantitatively reproduced by reconstruction of the muons in~AMANDA.

Monte Carlo simulation, based on the\break
 AMANDA-B4 reconstruction, predicts that\break
 AMANDA-B10 is a $10^4$~m$^2$ detector for TeV muons, with 2.5 degrees mean angular resolution per track\cite{christopher}. The effective area is less for atmospheric neutrinos, but in excess of $0.1\rm\,km^2$ for PeV neutrinos.

\section{\uppercase{Calibration on Atmospheric Neutrinos}}

Because of the novel technique, the collaboration has maintained 3 fully independent Monte Carlo programs simulating the signals, the detector medium and the detector itself. They quantitatively reproduce the response of the detector to cosmic ray muons: the trigger rate and the amplitude and arrival times of Cherenkov photons for each OM\cite{serap}. For reconstruction, 2 independent routines and 3 neural nets are available.

\begin{figure}[t]
\centering\leavevmode
\epsfxsize=7.5cm\epsffile{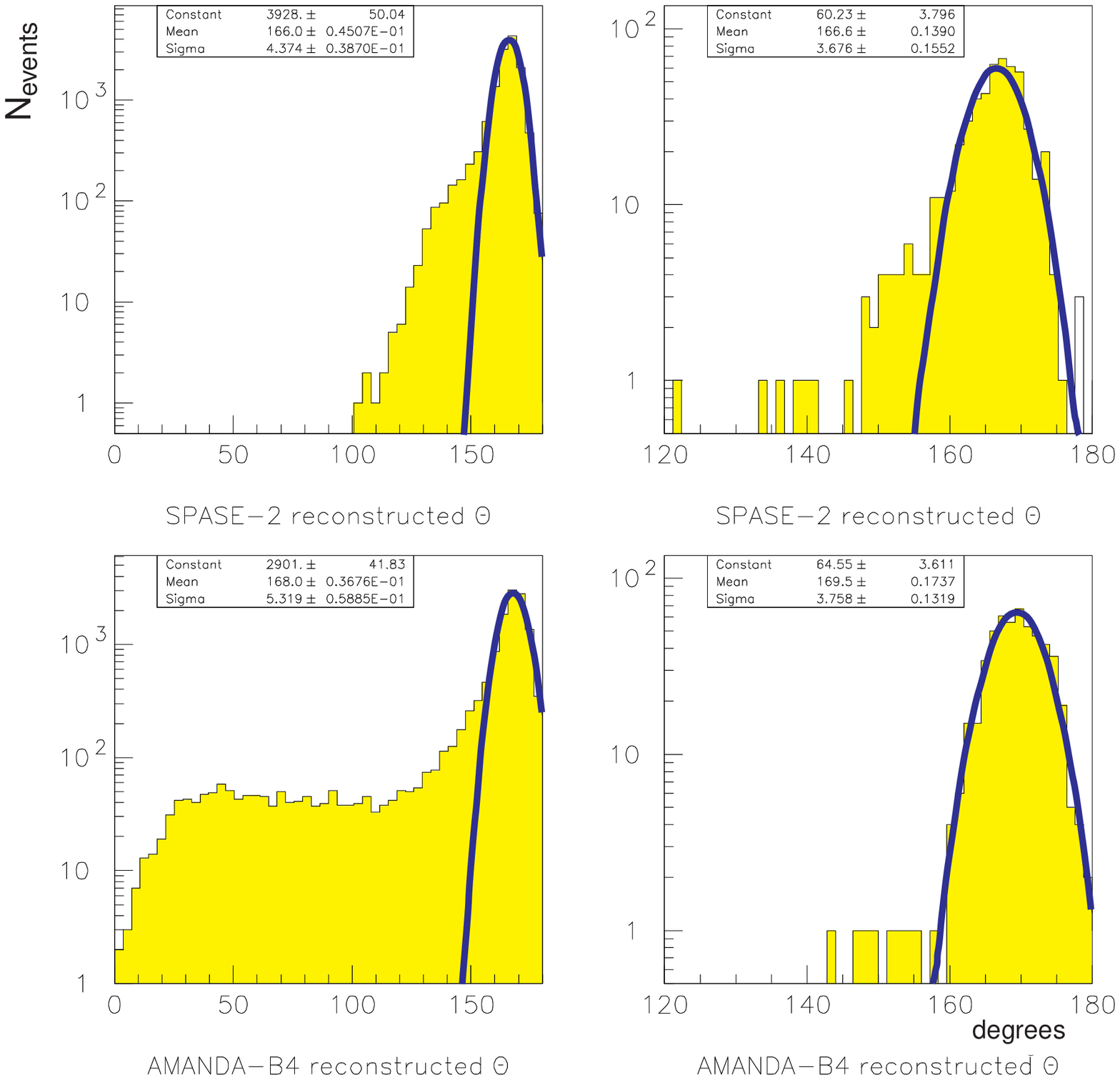}

\vspace{-2ex}

\caption{Zenith angle distributions of cosmic rays triggering AMANDA and the
surface air shower array SPASE. Reconstruction by AMANDA of underground muons
agrees with the reconstruction of the air shower direction using the scintillator array, and with Monte Carlo simulation. The events are selected requiring signals on 2 or more strings (left), and 5 or more direct photons (right).}
\end{figure}

Understanding the performance of the instrument near threshold requires a detailed calibration of the detector which is still in progress. Although  is not critical for operating the detector as a high energy neutrino telescope, it is for the detection of the flux of atmospheric neutrinos which falls sharply with energy. As a first calibration we have attempted to identify gold-plated events which are contained in the detector (within the instrumented volume and within $20^\circ$ of vertical) and which have a track-length in excess of 100\,m ($E_{\mu} > 20\,$GeV). Calculation of their rate is straightforward (see Table~3) except for the evaluation of the efficiency of the cut requiring 6 or more, direct photons with residuals in the
interval $[-15, +15]$\,ns. Monte Carlo simulation gives 5\%\cite{bouchta}. The narrow, long AMANDA-B4 detector (which constitutes the 4 inner strings of AMANDA-B10) thus achieves optimal efficiency for tracks which travel vertically upwards through the detector. Because of edge effects, the efficiency, which is of course a very strong function of detector size, is only a few percent after final cuts, even near the vertical direction. The bottom line is that we expect a few events per year satisfying the cuts imposed; see Table~3.  

\begin{table*}[t]
\def\arraystretch{1.15}\tabcolsep=1.25em
\caption{Predicted atmospheric neutrino rate for events with i) track-length in excess of 100\,m, ii) contained in the instrumented volume of the detector, iii) close to vertical direction, and iv) 6 or more direct hits. The results of AMANDA-B4 are contrasted with the anticipated rate for AMANDA-B10.}
\smallskip
\centering\leavevmode
\begin{tabular}{lcccc}
\hline
\noalign{\smallskip}
\multicolumn{5}{c}{\parbox{2in}{%
$\bullet$ close to vertical\\
$\bullet$ muon track $>100$ m}}\\[4mm]
\multicolumn{5}{c}{\# $\mu$'s$=375 \left[E_\mu\over20\rm\ GeV\right]^{-0.5}\rm\ (10^4\,m^2\,sr\ yr)^{-1}$}\\[2mm]
\hline\noalign{\smallskip}
&& \underline{\quad \vphantom{y}event rate\quad }&& \underline{\quad \vphantom{y}event rate\quad }\\[2mm]
radius& 35 m& 144 yr$^{-1}$ sr$^{-1}$& $>60$ m& 424 yr$^{-1}$ sr$^{-1}$\\[2mm]
\parbox[c]{.75in}{$\Delta\theta$ from\\ vertical}& 20$^\circ$& 70 yr$^{-1}$& $>45^\circ$& 1046 yr$^{-1}$\\[5mm]
\parbox[c]{.75in}{efficiency\\ $(N_{\rm dir} \ge 6)$}& 5\%& 3.5 yr$^{-1}$&
$>10\%$& 105 yr$^{-1}$\\[2mm]
rate& \multicolumn{2}{c}{found 2 in 6 months}& \multicolumn{2}{c}{$>0.3$ per day}\\[-4mm]
&\multicolumn{2}{c}{$\underbrace{\hspace{1.5in}}$}& \multicolumn{2}{c}{$\underbrace{\hspace{1.5in}}$}\\
& \multicolumn{2}{c}{80 OMs}& \multicolumn{2}{c}{300 OMs}\\
\hline
\end{tabular}
\end{table*}

We reconstructed 6 months of filtered\break
 AMANDA-B4 events subject to the conditions that 8 OMs report a signal in a time window of 2~microseconds. The two events, shown in Fig.\,10, satisfy the cuts outlined in Table~3. Their properties are summarized in Table~4. They have been used to study the capability of AMANDA to search for neutrinos resulting from the annihilation of dark matter particles gravitationally trapped at the center of the earth\cite{bouchta}.

\begin{figure}[t] 
\centering\leavevmode
\epsfysize=12cm\epsffile{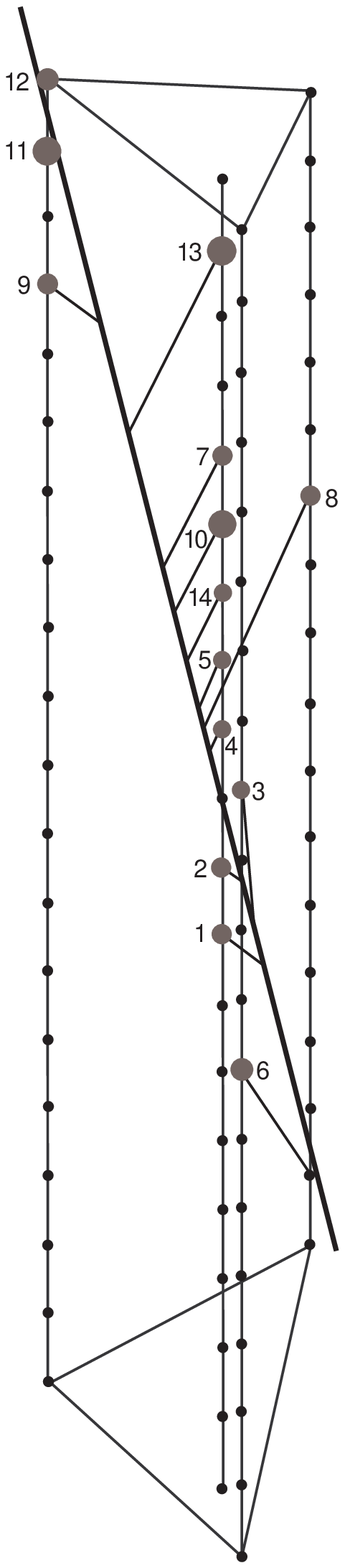}\hspace{1cm}
\epsfysize=12cm\epsffile{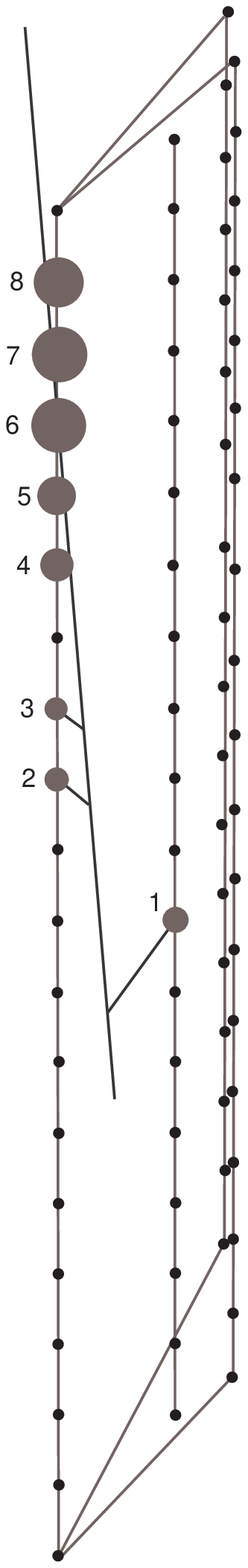}  

\caption{Events reconstructed as up-going satisfying the constrains of Table~3.}
\end{figure}

\begin{table}[t]
\caption{Characteristics of the two events
reconstructed as up-going muons.}
\label{tab:two_events}
\begin{center}
\begin{tabular} {lcc}
\hline
Event ID\# &4706879 & 8427905 \\
\hline
$\alpha$ [m/ns]& 0.19 & 0.37\\
Length [m] &295 & 182\\
Closest approach [m]&2.53  &1.23  \\
$\theta_{rec}$[$^\circ$] &14.1 & 4.6 \\
$\phi_{rec}$[$^\circ$]  &92.0  &  348.7\\
Likelihood/OM &  5.9 &  4.2 \\
OM multiplicity & 14 & 8  \\
String multiplicity & 4  &2  \\
\hline
\end{tabular}
\vspace{-4ex}
\end{center}
\end{table}

We conclude that tracks reconstructed as up-going are found at a rate consistent with the expectation that they are induced by atmospheric neutrinos. The event rates are too low to attempt a detailed calibration of the technique. The result is nevertheless encouraging because such events occur at the rate of about 1 per day in the full detector; see Table~3. Calibration of the full detector on atmospheric neutrinos should be feasible. This work is in progress and preliminary results based on 1 month of data are consistent with the performance of AMANDA as deduced from the AMANDA-B4 analysis.

\begin{table}[t]
\caption{Summary of the filtering of the 1997 data collected with the completed detector.}
\medskip
\centering\leavevmode
\tabcolsep=0em
\begin{tabular}{c}
\hline\noalign{\smallskip}
\epsfxsize=7.5cm\epsffile{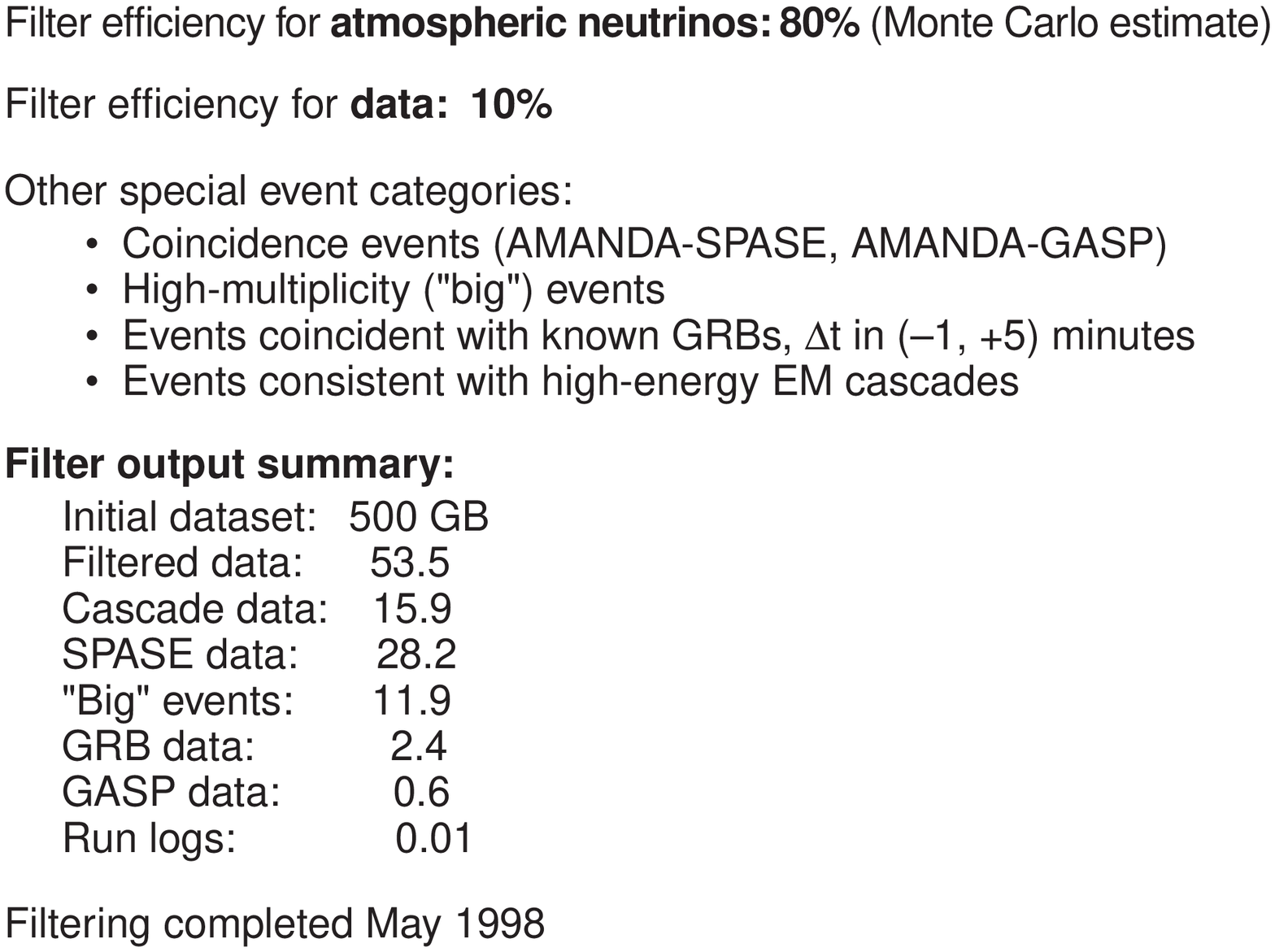}\\
\hline
\end{tabular}
\end{table}

\section{\uppercase{Data Analysis}}

Even with incomplete calibration, the detector can be operated as a high energy telescope. Events of PeV energy, predicted from such sources as gamma ray bursts and active galactic nuclei, are less challenging to identify than threshold atmospheric neutrinos. Our analysis procedure of the 1997 data collected with the completed AMANDA detector is sketched in Table~5. The 100\,Hz AMANDA-B10 trigger has generated a data set of 500 GigaBytes which has been reduced by a factor 10 by removing muon tracks that are clearly identified as down-going cosmic ray background events. This filtering required 1800~hours of Cray T3E time at NERSC/LBL. While it filters 65\% of the background, a Monte Carlo estimate is that 80\% of the atmospheric neutrino signal is retained. The filtered data set of only 500 GigaBytes can be analysed at the collaborating home institutions. Special filters also extracted events with the characteristics of large electromagnetic showers, events where more than 100 OMs report, events in coincidence with the SPASE air shower array and the GASP atmospheric Cherenkov telescope, and events within $(-1,+5)$~minutes of a gamma ray burst. Analysis of all categories of events is in progress.

AMANDA has also been operating as a burst detector of MeV neutrinos with, for instance, the capability of detecting galactic supernovae.

\section*{\uppercase{Acknowledgements}}

The AMANDA collaboration is indebted to the Polar Ice Coring Office and to Bruce Koci for the successful drilling operations, and to the National Science Foundation (USA), the Swedish National Research Council, the K.A.~Wallenberg Foundation and the Swedish Polar Research Secretariat. F.H.~is supported in part by the U.S.~Department of Energy under Grant No.~DE-FG02-95ER40896 and in part by the University of Wisconsin Research Committee with funds granted by the Wisconsin Alumni Research Foundation.

\begin{center}

{\bf $^\dagger$The AMANDA Collaboration\hfill}
\smallskip

R.C.~Bay,
D.~Chirkin,
Y.~He,
D.~Lowder,
P.~Miocinovic,
P.B.~Price,
W.~Rhode,
M.~Solarz,
K.~Woschnagg\\
{\it University of California, Berkeley, USA}\\

\smallskip

S.W.~Barwick,
J.~Booth,
J.~Kim,
P.C.~Mock,
R.~Porrata,
D.~Ross,
E.~Schneider,
W.~Wu,
G.~Yodh,
S.~Young\\
{\it University of California, Irvine, USA}\\

\smallskip

D. Cowen,
M.~Newcomer,
I.~Taboada\\
{\it University of Pennsylvania, USA}\\

\smallskip

M.~Carlson,
C.G.S.~Costa,
T.~DeYoung,
L.~Gray,
F.~Halzen,
R.~Hardtke,
G.~Hill,
J.~Jacobsen,
V.~Kandhadai,
A.~Karle,
I.~Liubarsky,
R.~Morse,
P.~Romenesko,
S.~Tilav\\
{\it University of Wisconsin, Madison, USA}\\

\smallskip

T.C.~Miller\\
{\it Bartol Research Institute,USA}\\

\smallskip

E.C.~Andr\'es,
P.~Askebjer,
L.~Bergstr\"om,
A.~Bouchta,
E.~Dalberg,
P.~Ekstr\"om,
A.~Goobar,
P.O.~Hulth,
J.~Rodriguez,
V.~Sorin,
C.~Walck\\
{\it Stockholm University, Sweden}\\

\smallskip

O.~Botner,
J.~Conrad,
A.~Hallgren,
C.~P.~de~los~Heros,
P.~Loaiza,
P.~Marciniewski,
H.~Rubinstein\\
{\it University of Uppsala, Sweden}\\

\smallskip

S.~Carius,
P.~Lindahl\\
{\it Kalmar University, Sweden}\\

\smallskip

A.~Biron,
S.~Hundertmark,
H.~Leich,
M.~Leuthold,
P.~Niessen,
T.~Schmidt,
U.~Schwendicke,
C.~Spiering,
P.~Steffen,
O.~Streicher,
T.~Thon,
C.H.~Wiebusch,
R.~Wischnewski\\
{\it DESY\,--\,Inst.\ for High Energy Physics, Germany}\\

\smallskip

W.~Chinowsky,
D.~Nygren,
G.~Przybylski,
G.Smoot,
R.~Stokstad\\
{\it Lawrence Berkeley National Laboratory, USA}\\

\smallskip

E.C.~Andr\'es
A.~Jones,
S.~Hart,
D.~Potter,
G.~Hill,
S.~Richter,
R.~Schwarz\\
{\it South Pole Winter-Overs, Antarctica}\\

\end{center}


\begin{thebibliography}{99}
\unskip

\frenchspacing

\bibitem{pr}
For a review, see T.\,K.\,Gaisser, F.\,Halzen and T.\,Stanev, {\it Phys.\ Rep.} {\bf
258}(3), 173 (1995); R.\,Gandhi, C.\,Quigg, M.\,H.\,Reno and I.\,Sarcevic, {\it
Astropart. Phys.}, {\bf 5}, 81 (1996).

\bibitem{science}
The AMANDA collaboration, {\it Science} {\bf 267}, 1147 (1995).

\bibitem{barwick}
S.\,W.\,Barwick {\it et al.}, {\it The status of the AMANDA high-energy neutrino detector}, in Proceedings of the 25th International Cosmic Ray Conference, Durban, South Africa (1997).

\bibitem{serap}
S.\,Tilav {\it et al.}, {\it First look at AMANDA-B data}, in Proceedings of
the 25th International Cosmic Ray Conference, Durban, South Africa (1997).

\bibitem{dublin}

S.\,W.\,Barwick {\it et al}, {\sl Proceedings of the 22nd International Cosmic
Ray Conference}, Dublin (Dublin Institute for Advanced Studies, 1991), Vol.~4,
p.~658.

\bibitem{christopher}
C.\,Wiebusch {\it et al.}, {\it Muon reconstruction with AMANDA-B}, in Proceedings of the 25th International Cosmic Ray Conference, Durban, South
Africa (1997).

\bibitem{miller}
T.\,Miller {\it et al.}, {\it Analysis of SPASE-AMANDA coincidence events}, in
Proceedings of the 25th International Cosmic Ray Conference, Durban, South
Africa (1997).

\bibitem{bouchta}
R.\,Bay {\it et al}, The AMANDA collaboration, Physics Reports {\bf 306}, to be published; A.\,Bouchta, University of Stockholm, PhD thesis (1998)


\end{thebibliography}
\end{document}